\documentclass[prd,aps,showpacs,nofootinbib,preprint,eqsecnum]{revtex4}
\usepackage[dvips]{graphicx}
\usepackage{color}
\usepackage{amssymb,amsmath}

\usepackage{amsxtra}
\usepackage{epsf}
\usepackage{amssymb}
\usepackage{enumerate}
\usepackage{hhline}
\usepackage{array}
\usepackage{tabularx}
\newcommand{\Eqn}[1]{&\hspace{-0.2em}#1\hspace{-0.2em}&}

\def\e{{\rm e}}

\begin{document}

\title{Periodic Cosmological Evolutions of Equation of State for 
Dark Energy}

\author{
Kazuharu Bamba$^{1, 3,}$\footnote{
E-mail address: bamba@kmi.nagoya-u.ac.jp}, 
Ujjal Debnath$^{2}$\footnote{
E-mail address: ujjal@iucaa.ernet.in, ujjaldebnath@yahoo.com}, 
Kuralay Yesmakhanova$^{3}$, 
Petr Tsyba$^{3}$, 
Gulgasyl Nugmanova$^{3}$ 
and 
Ratbay Myrzakulov$^{3,}$\footnote{
E-mail addresses: rmyrzakulov@csufresno.edu, rmyrzakulov@gmail.com}}
\affiliation{
$^1$Kobayashi-Maskawa Institute for the Origin of Particles and the
Universe,
Nagoya University, Nagoya 464-8602, Japan\\
$^2$Department of Mathematics, Bengal Engineering and Science University, Shibpur, Howrah-711 103, India\\
$^3$\textit{Eurasian International Center for Theoretical Physics,} \\ \textit{Eurasian National University, Astana 010008, Kazakhstan}}



%
%


\begin{abstract} 
We demonstrate two periodic or quasi-periodic generalizations of the Chaplygin gas (CG) type models to explain the origins of dark energy as well as 
dark matter by using the Weierstrass $\wp(t)$, $\sigma(t)$ and 
$\zeta(t)$ functions with two periods being infinite. 
If the universe can evolve periodically, 
a non-singular universe can be realized. 
Furthermore, 
we examine the cosmological evolution and nature of the equation of state (EoS) of dark energy in the Friedmann-Lema\^{i}tre-Robertson-Walker cosmology. 
It is explicitly illustrated that there exist three type models 
in which the universe always stays 
in the non-phantom (quintessence) phase, 
whereas it always evolves in the phantom phase, 
or the crossing of the phantom divide can be realized. 
The scalar fields and the corresponding potentials 
are also analyzed for different types of models. 
\end{abstract}

\pacs{
95.36.+x, 98.80.-k
}

\maketitle

\section{Introduction} 

Inflation in the early universe has been confirmed by 
the recent observations of 
cosmic microwave background (CMB) radiation~\cite{Spergel:2003cb, Spergel:2006hy, Komatsu:2008hk, Komatsu:2010fb}. 
In addition, the accelerated expansion of the current universe has also 
been suggested by recent 
observations, e.g., Type Ia Supernovae~\cite{Perlmutter:1998np, Riess:1998cb}, 
CMB radiation~\cite{Spergel:2003cb, Spergel:2006hy, Komatsu:2008hk, Komatsu:2010fb}, 
the large scale structure LSS~\cite{Tegmark:2003ud, Seljak:2004xh}, 
baryon acoustic oscillations (BAO)~\cite{Eisenstein:2005su}, 
and weak lensing~\cite{Jain:2003tba}. 
To explain such a cosmic acceleration, one provides 
the existence of so-called dark energy in the framework of 
general relativity (for reviews, see, e.g.,~\cite{
Copeland:2006wr, Durrer:2007re, Durrer:2008in, Cai:2009zp, Tsujikawa:2010sc, 
Book-Amendola-Tsujikawa, Li:2011sd, Bamba:2012cp}), 
or one supposes that gravity is modified on the large scale 
(for reviews, see, e.g.,~\cite{Nojiri:2010wj, Nojiri:2006ri, 
Sotiriou:2008rp, Book-Capozziello-Faraoni, Capozziello:2011et, 
DeFelice:2010aj, Clifton:2011jh, Capozziello:2012hm}). 

In the expansion history of the universe, there exist two 
singularities. One is a Big Bang singularity. 
The other is the finite-time future singularities~\cite{Caldwell:2003vq, McInnes:2001zw, Nojiri:2003vn, Nojiri:2003ag, Faraoni:2001tq, GonzalezDiaz:2003rf, Elizalde:2004mq, Singh:2003vx, Csaki:2004ha, Wu:2004ex, Nesseris:2004uj, Sami:2003xv, Stefancic:2003rc, Chimento:2003qy, Hao:2004ky, Elizalde:2005ju, Dabrowski:2004hx, Lobo:2005us, Cai:2005ie, Aref'eva:2005fu, Lu:2005qy, Godlowski:2005tw, Sola:2005et, Guberina:2005mp, Shtanov:2002ek, Barrow:2004xh, Nojiri:2004ip, Nojiri:2004pf, Nojiri:2005sr, Cotsakis:2004ih, Dabrowski:2004bz, FernandezJambrina:2004yy, FernandezJambrina:2008dt, Barrow:2004he, Stefancic:2004kb, Cattoen:2005dx, Tretyakov:2005en, Balcerzak:2006ac, Sami:2006wj, BouhmadiLopez:2006fu, Yurov:2007tw, Koivisto:2008xfa, Barrow:2009df, BouhmadiLopez:2009jk, Nojiri:2005sx, 
Nojiri:2008fk, Bamba:2008ut, Bamba:2009uf}, which occurs at the last stage of the universe filled with dark energy, 
or a Big Crunch singularity. 
To avoid these singularities, various cosmological scenarios have been 
proposed, e.g., 
the cyclic universe~\cite{Steinhardt:2001st, Steinhardt:2001vw, Khoury:2001zk, Steinhardt:2002kw, Steinhardt:2002ih, Khoury:2003rt, Steinhardt:2006bf, Saaidi:2012qp, Nojiri:2011kd, Cai:2011bs, Sahni:2012er} (for a reference in a different context, see~\cite{Chung:2001ka}), 
the ekpyrotic scenario~\cite{Khoury:2001wf, Khoury:2001iy, Donagi:2001fs, Khoury:2001zk}, 
and the bouncing universe~\cite{Page:1984qt, Peter:2001fy, Peter:2002cn, Shtanov:2002mb, Biswas:2005qr, Cai:2007qw, Creminelli:2007aq, Piao:2009ku, Zhang:2010bb, Piao:2010cf, Liu:2012gu, Novello:2008ra}. 
Furthermore, related to the cyclic universe, 
the (trefoil and figure-eight) knot universe has been investigated 
in Ref.~\cite{Myrzakulov:2010tc, RM-Knot-Universes, Esmakhanova:2011ar, Jamil:2012jd}. 
In addition, motivated by the studies on the role of applying the Weierstrass $\wp(t)$, $\zeta(t)$ and $\sigma(t)$-functions and the Jacobian elliptic functions to astrophysics and cosmology~\cite{Gibbons:2011rh, Bochicchio:2011wx, Dimitrov:2001sy, D'Ambroise:2009ax, BouhmadiLopez:2002qz}, 
the equation of state (EoS) for the cyclic universes 
in the homogeneous and isotropic Friedmann-Lema\^{i}tre-Robertson-Walker (FLRW) spacetime has been reconstructed by using 
the Weierstrass and Jacobian elliptic functions in 
Ref.~\cite{Bamba:2012gq}. 

In this paper, based on the reconstruction method in Refs.~\cite{Nojiri:2010wj, Nojiri:2006ri, Nojiri:2005sx, Nojiri:2005sr, Stefancic:2004kb}, 
with the Weierstrass $\wp(t)$-function, 
we examine the cosmological evolution of the EoS for dark energy 
in FLRW cosmology. 
In particular, it is shown that two periodic generalized Chaplygin gas (GCG) type models for dark energy can be reconstructed. 
To account for the origins of dark energy as well as dark matter with a fluid, 
the original CG~\cite{Kamenshchik}, GCG~\cite{Bento:2002ps} 
and the modified CG (MGC)~\cite{Benaoum:2002zs, Sharif:2012cu} have been explored. 
We mention that 
the reconstruction of periodic cosmologies have widely been studied. 
Especially, the reconstruction of periodic EoS has been investigated, 
e.g., in Ref.~\cite{SaezGomez:2008qe}. 
In this reference, 
with an inhomogeneous EoS for dark energy fluid, 
it has been demonstrated that an oscillating universe can occur. 
Also, the Hubble parameter with a periodic behavior 
can realize both inflation in the early universe and the late-time cosmic 
acceleration under the same mechanism in a unified manner. 
In addition, it has been verified that 
a coupling between dark energy fluid, which has 
a homogeneous and constant EoS, and matter, 
can present a periodic behavior of the universe. 
Furthermore, as several theoretical issues in the universe with 
its oscillatory behavior, the phantom phase and finite-time future 
singularities have been investigated. 
A scalar-tensor description of the oscillating universe 
has also been explored. 
As stated, there exist various theoretical subjects in the periodic 
cosmological evolution of the universe. 
The essential property of the Weierstrass functions 
is to have two periods $m_1$ and $m_2$. 
Hence, the periodical and quasiperiodical models which we show in Sec.~3
are periodical or quasiperiodical in terms of the energy density $\rho$. 
In addition, these models with the periods $m_1$ and $m_2$ being infinite 
are reduced to the Chaplygin gas models 
(as is seen from the formulae in Eq.~(\ref{eq:3-G-ADD-02})). 
Thus, the reconstruction procedure of these models corresponds to 
two periodic or quasi periodical generalizations of the CG models. 
This justifies the use of the Weierstrass functions in cosmological models. 
Furthermore, the models given in Sec.~4 are periodic on the cosmic time $t$ 
(which is a dimensionless quantity in our analysis), although 
the models in Sec.~3 are periodic/quasiperiodic on $\rho$. 
The periodicity of the cosmological evolution comes from the 
periodic nature of the Weierstrass functions. 
Also, the periodic/quasiperiodic models in Sec.~3 are singular at $\rho=0$, 
whereas those in Sec.~4 are singular at $t=0$. 
If the periodic evolution of the universe can be realized, 
various scenarios to avoid cosmological singularities can be constructed. 
This is the important cosmological motivation to obtain such periodical 
solutions. 
Moreover, we explicitly demonstrated that there exist three type models 
in which (i) the universe always stays 
in the non-phantom (quintessence) phase, 
(ii) it always evolves in the phantom phase, 
and (iii) the crossing of the phantom divide can be realized. 
It has recently been shown that these three cases have also been realized in non-local gravity~\cite{Bamba:2012ky}. 
It is also interesting to remark that 
according to the analysis of recent cosmological observational data, 
in the past the crossing of the phantom divide occurred~\cite{Alam:2004jy, Alam:2006kj, Nesseris:2006er, Wu:2006bb, Jassal:2006gf}. 
We use the units of 
the gravitational constant 
$8 \pi G = c =1$ with $G$ and $c$ being the gravitational constant and the seed of light. 

The paper is organized as follows. 
In Sec.\ 2, we show the basic equations in the FLRW background and briefly give the Chaplygin gas type models. 
In Sec.\ 3, we study periodical and quasi-periodical GCG type models. 
In Sec.\ 4, we demonstrate other two periodical FLRW models. 
Finally, several conclusions are presented in Sec.\ 5. 

\section{Brief review of the CG type models}

In this section, we  
briefly explain the significant features of the 
CG type models for the spatially flat homogeneous and 
isotropic FLRW universe. 
The action describing general relativity and matter is given by
$
S=\int\sqrt{-g}d^4x (R+L_m)
$, 
where $R$ is the scalar curvature and $L_m$ is the matter Lagrangian. 
We take the flat 
FLRW spacetime with the metric, 
$
ds^2=-dt^2+a^2(t)\left(dr^2+r^2d\Omega^2\right)
$. 
Here, $a(t)$ is the scale factor and 
$d \Omega^2$ is the metric of 2-dimensional sphere with unit radius. 
We note that in this paper, time ($t$) is considered to be a dimensionless 
quantity. 
In the flat FLRW background, 
from the above action we obtain the gravitational field equations
\begin{eqnarray}
\left( \frac{\dot{a}}{a} \right)^2 \Eqn{=} \frac{\rho}{3}\,, 
\label{FE1}\\
\frac{\ddot{a}}{a} \Eqn{=} - \frac{\rho + 3p}{6}\,. 
\label{FE2}
\end{eqnarray} 
Here, the Hubble parameter is defined by $H \equiv \dot{a}/a$ and a dot denotes the time derivative of $\partial/\partial t$. 
By using these equations, we have the expressions of the energy density 
$\rho=3\left(\dot{a}/a\right)^2$ and pressure $p=-2\left(\ddot{a}/a\right) 
- \left(\dot{a}/a\right)^2$. 

Next, we explore the CG type models~\cite{Kamenshchik, Bento:2002ps, Benaoum:2002zs, Sharif:2012cu}. 
The GCG model has been constructed in order to account for both the origins of dark energy and dark matter with using a single fluid. 
The equation of state (EoS) of the GCG is given by~\cite{Bento:2002ps} 
\begin{equation} 
p = -\frac{C_1}{\rho^{\alpha}}\,, 
\label{eq:2-B-1} 
\end{equation}
where $C_1 (>0)$ is a positive constant and $\alpha$ is a constant. 
If we take $\alpha=1$, 
Eq.~(\ref{eq:2-B-1}) describes 
the original CG model~\cite{Kamenshchik}. 
{}From Eq.~(\ref{eq:2-B-1}) and the continuity equation 
$\dot{\rho}=-3H(\rho+p)$, 
we obtain 
\begin{equation} 
\rho = \left[C_1 + \frac{C_2}{a^{3\left(1+\alpha\right)}} 
\right]^{1/\left(1+\alpha\right)}\,, 
\label{eq:IIIB-add-04} \\ 
\end{equation}
where $C_2$ is a constant of integration. 
{}From Eq.~(\ref{eq:IIIB-add-04}), we find the asymptotic behaviors of 
$\rho$ that 
in the early universe $a \ll 1$, 
$\rho \sim C_2 a^{-3}$, 
whereas 
in the late universe $a \gg 1$,  
$\rho \sim C_1^{1/\left(1+\alpha\right)}$. 
Thus, 
in the early universe, the energy density behaves as $\rho \propto a^{-3}$, 
which is the same as  non-relativistic matter such as dark matter. 
On the other hand, in the late universe the energy density becomes a constant 
as $\rho \to C_1^{1/\left(1+\alpha\right)}$. This means that it can play a role of dark energy. 
As a consequence, 
the GCG model 
can explain the origin of dark energy as well as dark matter simultaneously. 

In addition, the MCG has been proposed in Refs.~\cite{Benaoum:2002zs, Sharif:2012cu}. 
The EoS is given by 
\begin{equation}
p=C_3\rho-\frac{C_4}{\rho^{\alpha}}\,, 
\label{eq:2-B-2} 
\end{equation}
where $C_3$ and $C_4 (>0)$ are constants.

\section{Periodical and quasi-periodical GCG type models}

In this section, we examine the periodical and quasi-periodical 
GCG type models by using the Weierstrass functions, the so-called MG -- $i$ 
models\footnote{Note that the meaning of the so-called Myrzakulov Gas (MG) -- $i$ ($i$ = XII, XIII, XIV, XV, XVI, XVII, XVIII, XIX, XX, XXXIII, XXI, XXII, XXII, XXIII, XXVII) is the model of some gases/fluid, 
which is the notation used in Refs.~\cite{Myrzakulov:2010tc, RM-Knot-Universes, Esmakhanova:2011ar, Jamil:2012jd}.}. 
Properties of elliptic functions inform 
us that the MG -- XXV, MG -- XXVI and MG -- XXIV models (as the MG -- XXIII and MG -- XXVI models) are some generalizations of the CG type models 
due to the following degenerate cases of some elliptic and related functions as $m_1=\infty$ and $m_2=\infty$, 
where $m_1$ and $m_2$ are two periods~\cite{Book-elliptic-functions}:
\begin{equation}
\sigma(x)=x\,, \quad 
\zeta(x)=x^{-1}\,, \quad 
\wp(x)=x^{-2}\,, \quad 
am(x)=x\,. 
\label{eq:3-G-ADD-02}
\end{equation}
Here, $g_2=g_3=0$ and $e_1=e_2=e_3=0$, where $g_2$ and $g_3$ are the Weierstrass invariants. 

The physical motivation to examine the series of the MG -- $i$ gas 
is as follows. 
These models can realize the cosmological evolution 
of the GCG type models with 
the periodical and quasi-periodical behaviors, 
which depends on the models. 
These models are expressed with the Weierstrass functions 
and hence various behaviors of the cosmic expansion history 
with periodicity and/or quasi-periodicity 
can be realized. 
Thus, these models can present novel cosmological scenarios without 
a Big Bang singularity in the early universe and 
 the finite-time future singularities or a Big Crunch singularity, 
such as the cyclic universe, 
the ekpyrotic scenario and the bouncing universe. 

\subsection{Periodical generalizations}

\subsubsection{MG -- XXI model}

One of the most interesting examples of gases 
is the MG -- XXI model 
which has the following EoS~\cite{Myrzakulov:2010tc, RM-Knot-Universes, Esmakhanova:2011ar, Jamil:2012jd}
\begin{equation}
p=-B[\wp(\rho)]^{0.5}\,, 
\label{eq:3ADD-001}
\end{equation}
where $B(>0)$ is a positive constant. 
By using the degenerate case of the function $\wp(\rho)$ 
in~(\ref{eq:3-G-ADD-02}), we can show that 
the well-known CG model~\cite{Kamenshchik}
$
p=-B/\rho
$, which is equal to Eq.~(\ref{eq:IIIB-add-04}) with $B = C_1$ and 
$\alpha = 1$, 
is particular case of the MG -- XXI model in Eq.~(\ref{eq:3ADD-001}). 
The parameter of the EoS $\omega$ 
for our model is given by 
\begin{equation}
\omega \equiv \frac{p}{\rho} = -\frac{B\sqrt{\wp(\rho)}}{\rho}\,. 
\label{eq:3ADD-002}
\end{equation}

%
%

%
%

\subsubsection{MG -- XXII model}

One of the two periodical generalizations of the GCG 
is given by~\cite{Myrzakulov:2010tc, RM-Knot-Universes, Esmakhanova:2011ar, Jamil:2012jd}
\begin{equation}
 p=-B[\wp(\rho)]^{0.5\alpha}\,.
\end{equation}
In fact, 
its degenerate case is the GCG~\cite{Bento:2002ps} 
$
p=-B/\rho^{\alpha} 
$, which is equivalent to Eq.~(\ref{eq:2-B-1}) with $B = C_1$. 
For this model, the parameter of the EoS looks like 
\begin{equation}
\omega = -\frac{B[\wp(\rho)]^{0.5\alpha}}{\rho}\,. 
\end{equation}

\subsubsection{MG -- XXIII model}

Next, 
we present 
one of the two periodical generalizations of the MCG. 
Its EoS reads~\cite{Myrzakulov:2010tc, RM-Knot-Universes, Esmakhanova:2011ar, Jamil:2012jd}
\begin{equation}
p=A\rho-B[\wp(\rho)]^{0.5\alpha}\,, 
\end{equation}
where $A$ is a constant. 
The corresponding parameter of the EoS is 
\begin{equation}
\omega = A-\frac{B[\wp(\rho)]^{0.5\alpha}}{\rho}\,. 
\end{equation}

\subsubsection{MG -- XXIV model}

We now give a more general form of two periodical generalizations of the MCG. Its EoS is described as 
\begin{equation}
p=A[\wp(\rho)]^{-0.5}-B[\wp(\rho)]^{0.5\alpha}\,.
\end{equation}
The parameter of the EoS for the model is written by 
\begin{equation}
\omega = \frac{A[\wp(\rho)]^{-0.5}}{\rho}-\frac{B[\wp(\rho)]^{0.5\alpha}}{\rho}\,. 
\label{eq:3A4-add-1}
\end{equation}
Again, by using the degenerate properties of the elliptic functions, 
we can demonstrate that this model is reduced to MCG. 

%
%

%
%

\subsection{Quasi-periodical generalizations}

In the preceding subsection, we have considered 
two periodical generalizations of CG type models. 
In this subsection, we study quasi-periodical models. 

\subsubsection{MG -- XXV model} 
One of the quasi-periodical models, the so-called 
MG -- XXV model, 
is given by 
\begin{equation}
p=A\sigma(\rho)-\frac{B}{[\sigma(\rho)]^{\alpha}}\,,
\label{eq:3-E-ADD-01}
\end{equation}
where $A$ and $B$ are constants and $\sigma(\rho)$ is the Weierstrass 
$\sigma$-function. 
As the $\sigma$-function degenerates according to equations in 
(\ref{eq:3-G-ADD-02}), in this case Eq.~(\ref{eq:3-E-ADD-01}) becomes 
the MCG in Eq.~(\ref{eq:2-B-2}). 
The corresponding parameter of the EoS is expressed as 
\begin{equation}
\omega = \frac{A\sigma(\rho)}{\rho}-\frac{B}{\rho[\sigma(\rho)]^{\alpha}}
\,. 
\label{eq:3BADD-001}
\end{equation}

%
%

%
%

\subsubsection{MG -- XXVI model}

Our next quasi-periodical model is given by 
\begin{equation}
p=\frac{A}{\zeta(\rho)}-B[\zeta(\rho)]^{\alpha}\,,
\label{eq:3-F-ADD-01}
\end{equation}
where $\zeta(x)$ is the Weierstrass $\zeta(x)$-function. 
It is the MG -- XXVI model. 
For the degenerate case in Eq.~(\ref{eq:3-G-ADD-02}), 
this model is also reduced to the MCG. 
Its parameter of the EoS takes the form 
\begin{equation}
\omega = \frac{A}{\rho\zeta(\rho)}-\frac{B[\zeta(\rho)]^{\alpha}}{\rho}
\,. 
\label{eq:3B2-add-1}
\end{equation}

%
%

%
%

\subsubsection{MG -- XXVII model}

We explore the MG -- XXVII model. For this model, the EoS reads
\begin{equation}
p=A am(\rho)-B[am(\rho)]^{-\alpha}\,,
\label{eq:3-G-ADD-01}
\end{equation}
where $am(x)$ is the Jacobi amplitude ($am(x)$) function and 
$\alpha$ is a constant. 
In case of the degeneration in Eq.~(\ref{eq:3-G-ADD-02}), 
this model recovers the MCG. 
The parameter of the EoS is written by 
\begin{equation}
\omega= \frac{A am(\rho)}{\rho} 
- \frac{B[am(\rho)]^{-\alpha}}{\rho}\,,
\label{eq:3B3-add-1}
\end{equation}

%
%

It is significant to emphasize that 
(a) if we substitute $\sigma(t)$ in Eq.~(\ref{eq:3-G-ADD-02}) into 
Eq.~(\ref{eq:3-E-ADD-01}), 
(b) if we use $\zeta(t)$ in Eq.~(\ref{eq:3-G-ADD-02}) and 
Eq.~(\ref{eq:3-F-ADD-01}), 
(c) if we combine $\wp(t)$ in Eq.~(\ref{eq:3-G-ADD-02}) with 
Eq.~(\ref{eq:3-G-ADD-01}), 
then 
we obtain the MGC in Eq.~(\ref{eq:2-B-2}). 
As a result, in the limit of $m_1=\infty$ and $m_2=\infty$, 
the MG -- XXV, MG -- XXVI and MG -- XXIV models are reduced to 
the MCG~\cite{Benaoum:2002zs, Sharif:2012cu}. 
This point is the most important and novel observataion in this work. 

%
%

In the limit of the small energy density $\rho \to 0$ as well as 
$\rho \to \infty$, the behaviors of the EoS for the universe 
in the MG -- XXI, MG -- XXIV, MG -- XXV, MG -- XXVI and MG -- XXVII models 
asymptotically approach those in the CG model. 
On the other hand, in the middle regime of $\rho$, 
since the EoS for the universe in 
the MG -- XXI, MG -- XXIV, MG -- XXV, MG -- XXVI and MG -- XXVII models 
is described by using elliptic functions with a periodic or quasi-periodic 
property, 
the EoS for the universe expresses also 
periodic or quasi-periodic behaviors. 

{}From the above considerations, 
the cosmological evolution of the universe is described as follows. 
First, the energy responsible for inflation would be released to 
radiation (i.e., relativistic matter) through a reheating process
and the universe enter the radiation dominated stage. 
Here, the concrete mechanism for both inflation and the reheating 
stage is not specified. 
After that, as the universe expands, its temperature decreases with 
proportional to $a^{-1}$, and the matter (i.e., non-relativistic matter) 
dominated stage appears. This can be seen in our models in the limit of 
$\rho \to \infty$, namely $\omega$ asymptotically approaches zero, which 
corresponds to the EoS of the dust. 
Finally, the universe becomes the dark energy dominated stage. 
This can also be understood in the limit of 
$\rho \to 0$, where $\omega <-1/3$. 
Thus, it is considered that the cosmological evolution of 
the universe can be realized in our models.

%
%

\section{Other two periodical FLRW models}

The EoS for dark energy is one of the most significant cosmological 
quantities. In this paper, we concentrate on the evolution of the EoS 
for dark energy. 
In the FLRW spacetime, 
the effective EoS for the universe 
is given by~\cite{Nojiri:2010wj, Nojiri:2006ri} 
$
\omega_{\mathrm{eff}} \equiv p_{\mathrm{eff}}/\rho_{\mathrm{eff}} = 
-1 - 2\dot{H}/\left(3H^2\right)
$. Here, $\rho_{\mathrm{eff}}$ and $p_{\mathrm{eff}}$ can be 
considered to the total energy density and pressure of the universe, 
respectively. 
Since we examine the dark energy dominated stage, the energy density 
$\rho_{\mathrm{DE}}$
and pressure $p_{\mathrm{DE}}$ of dark energy can be regarded as 
$\rho_{\mathrm{DE}} \approx \rho_{\mathrm{eff}}$ and 
$p_{\mathrm{DE}} \approx p_{\mathrm{eff}}$. 
As a result, we find $\omega_{\mathrm{DE}} \approx \omega_{\mathrm{eff}}$. 

In addition, we represent $\rho_{\mathrm{DE}}$ and $p_{\mathrm{DE}}$ 
as $\rho$ in Eq.~(\ref{FE2}) and $p$ in Eq.~(\ref{FE1}), respectively. 
In the non-phantom (quintessence) phase, 
$\dot{H} < 0$ and hence $\omega_\mathrm{eff} >-1$
, which is the non-phantom (quintessence) phase, 
while in the phantom phase, 
$\dot{H} > 0$ and therefore $\omega_\mathrm{eff} <-1$. 
If $\dot{H} = 0$, $\omega_\mathrm{eff} =-1$, which is the case that 
dark energy is the cosmological constant. 

As a qualitative criterion to constrain the models, 
we examine the evolution of the EoS $\omega$ of a fluid corresponding to 
dark energy. 
If $\omega$ is always less than $-1$, the universe stays in the phantom phase 
in all the cosmic evolution history. This case is clearly inconsistent with 
the standard cosmological evolution and hence it can be ruled out. 
On the other hand, if $\omega$ is always larger than $-1$ or it crosses 
the line of $-1$, these cases are not ruled out, namely these models 
may have the possibility to realize the standard evolution history 
of the universe. 

In this section, we study new two periodical FLRW models. 
These models are expressed by using 
the Weierstrass $\wp(t)$-function, which 
as well known satisfies 
the following ordinary differential equations~\cite{Book-elliptic-functions} 
\begin{eqnarray}
\dot{\wp}^{2}(t)\Eqn{=}4\wp^3(t)-g_2\wp(t)-g_3\,,\\
\ddot{\wp}(t)\Eqn{=}6\wp^2(t)-0.5g_2\,,\\
\dddot{\wp}(t)\Eqn{=}12\wp(t)\dot{\wp}(t)\,,\\
\ddddot{\wp}(t)\Eqn{=}120\wp^3(t)-18g_2\wp(t)-12g_3\,, 
\end{eqnarray}
where $\dot{\wp}(t) = d\wp(t)/dt$ and so on. 
In what follows, by using the reconstruction method in Refs.~\cite{Nojiri:2010wj, Nojiri:2006ri, Nojiri:2005sx, Nojiri:2005sr, Stefancic:2004kb}, 
and the Weierstrass $\wp(t)$-function, 
for ten models [the MG -- $i$ models ($i$ = XII, XIII, XIV, XV, XVI, XVII, XVIII, XIX, XX, XXXIII)], 
we reconstruct the EoS for dark energy and 
explore its cosmological evolution in FLRW cosmology.

\subsection{MG -- XII model}

We suppose that 
the Hubble parameter is represented as 
\begin{equation}
H=\wp(t)\,.
\label{FIG1_1}
\end{equation}
{}From this expression, the scale factor becomes 
\begin{equation}
a(t)=a_0\e^{-\zeta(t)}\,,
\label{FIG1_2}
\end{equation}
where $a_0 (>0)$ is a positive constant and 
$\zeta(t)$ is the Weierstrass $\zeta(t)$-function. 
Then, 
Eq.~(\ref{FIG1_1}) and 
the gravitational field 
equations (\ref{FE1}) and (\ref{FE2}) lead to 
the parametric EoS 
\begin{eqnarray}
p\Eqn{=}-2\sqrt{4\wp^3(t)-g_2\wp(t)-g_3}-3\wp^2(t)\,,
\label{FIG1_3} \\ 
\rho\Eqn{=}3\wp^2(t)\,.
\label{FIG1_4}
\end{eqnarray} 
By using Eqs.~(\ref{FIG1_3}) and (\ref{FIG1_4}), we see that 
the EoS parameter is written as 
\begin{equation}
\omega = \frac{p}{\rho}
=-1-\frac{2\sqrt{4\wp^3(t)-g_2\wp(t)-g_3}}{3\wp^2(t)}\,.
\label{FIG1_5}
\end{equation}

In order to describe our models in terms of the scalar field theory, we 
introduce a scalar field $\phi$ and its self-interacting 
potential $V(\phi)$. 
The Lagrangian for the scalar field theory is given by (see, e.g.,~\cite{Nojiri:2010wj, Nojiri:2006ri})
\begin{equation}
L_{\phi}=0.5\dot{\phi}^2-V(\phi)\,.
\end{equation}
Thus, this scalar is corresponding to a phantom one with 
$\omega < -1$, which can be seen in Eq.~(\ref{FIG1_5}). 
The energy momentum tensor of 
the scalar field $\phi(t)$ 
is identical to a fluid with the energy density
$\rho_{\phi}$ and pressure $p_{\phi}$ given by 
\begin{eqnarray}
\rho_{\phi}\Eqn{=}-0.5\dot{\phi}^2+V(\phi)=\rho\,, \\
p_{\phi}\Eqn{=}-0.5\dot{\phi}^2-V(\phi)=p\,.
\end{eqnarray}
By using these expressions, we find 
\begin{eqnarray}
-\dot{\phi}^2\Eqn{=}\rho+p\,, 
\label{SF}\\
V(\phi)\Eqn{=}0.5(\rho-p)\,.
\label{SIP}
\end{eqnarray}

In addition, it follows from Eqs.~(\ref{SF}) and (\ref{SIP}) that 
the scalar field $\phi$ and self-interacting
potential $V(\phi)$ are written as 
\begin{eqnarray}
\phi\Eqn{=}i\sqrt{2}\int\sqrt{\sqrt{4\wp^3(t)-g_2\wp(t)-g_3}}dt\,, \\
V\Eqn{=}3\wp^2\left(t\right)+\sqrt{4\wp^3\left(t\right)-g_2\wp\left(t\right)-g_3}\,.
\end{eqnarray}

\begin{figure}[h!]
  \centering
  \includegraphics[]{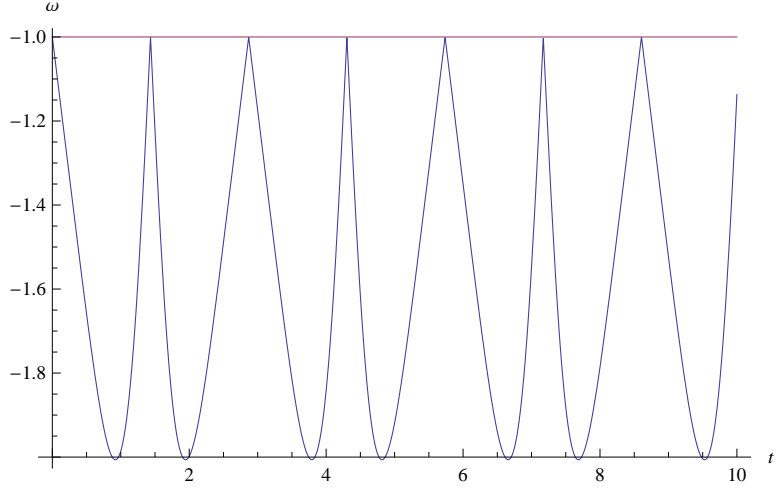}
  \caption{The EoS $\omega$ in Eq.~(\ref{FIG1_5}) as a function of $t$ for 
$\wp\left(t,1,1\right)$, i.e., the model parameters of the Weierstrass invariants of $g_2 =1$ and $g_3 =1$. 
The line of $\omega = -1$ is also plotted.}
  \label{fig:Example-1_5}
\end{figure}

In Fig.~\ref{fig:Example-1_5}, we show the cosmological evolution of 
EoS $\omega$ as a function of $t$ for 
$\wp\left(t,1,1\right)$, i.e., the model parameters of the Weierstrass invariants of $g_2 =1$ and $g_3 =1$. 
From Fig.~\ref{fig:Example-1_5}, we see that the universe always stays 
in the phantom phase ($\omega < -1$). 
Hence, this model is ruled out. 
Furthermore, we find the two periodic 
oscillatory behavior of $\omega$.

%
%

We remark a point in terms of the expression of $V$. 
In the above procedure, 
first the form of the scale factor $a=a(t)$ or the Hubble parameter $H=H(t)$ is supposed. 
Next, from this form we obtain the pressure $p=p(t)$ and the energy density $\rho=\rho(t)$. 
On the other hand, in the description of the scalar field theory, 
the scalar field 
$\phi = \phi(t)$ and its potential $V=V(\phi)$ are expressed with $p=p(t)$ and 
$\rho=\rho(t)$. 
Accordingly, $V=V(\phi)=V(\phi(t))$. This means that $V$ is written as 
a function of cosmic time $t$. 
Thus, in principle, if $a$, $H$, $p$ and $\rho$ are inversely solved 
in terms of $t$, $V$ can also be described by the expressions of $a$, $H$, $p$ 
and $\rho$. 

\subsection{MG -- XIII model}

We express the Hubble parameter as 
\begin{equation}
H=\dot{\wp}(t)\,.
\label{FIG2_1}
\end{equation}
{}From this expression, the scale factor is given by
\begin{equation}
a(t)=a_0\e^{\wp(t)}\,.
\label{FIG2_2}
\end{equation}
By combining Eq.~(\ref{FIG2_1}) with the gravitational field equations 
(\ref{FE1}) and (\ref{FE2}), 
we find that 
the parametric EoS is written as 
\begin{eqnarray}
p\Eqn{=}-12\wp^3(t)-12\wp^2(t)+3g_2\wp(t)+g_2+3g_3\,,
\label{FIG2_3}\\
\rho\Eqn{=}3\left(4\wp^3(t)-g_2\wp(t)-g_3\right)\,.
\label{FIG2_4}
\end{eqnarray}
It follows from Eqs.~(\ref{FIG2_3}) and (\ref{FIG2_4}) that 
the EoS parameter is given by
\begin{equation}
\omega=-\frac{12\wp^3(t)+12\wp^2(t)-3g_2\wp(t)-g_2-3g_3}{3\left(4\wp^3(t)-g_2\wp(t)-g_3\right)}\,.
\label{FIG2_5}
\end{equation}
By using Eqs.~(\ref{SF}) and (\ref{SIP}), 
we 
obtain the expressions of the 
scalar field $\phi$ and self-interacting potential $V(\phi)$
\begin{eqnarray}
\phi\Eqn{=}\int\sqrt{-12\wp^2\left(t\right)+g_2}dt\,, \\
V\Eqn{=}12\wp^3\left(t\right)+6\wp^2\left(t\right)-3g_2\wp\left(t\right)-3g_3-0.5g_2\,.
\end{eqnarray}

\begin{figure}[h!]
  \centering
  \includegraphics[]{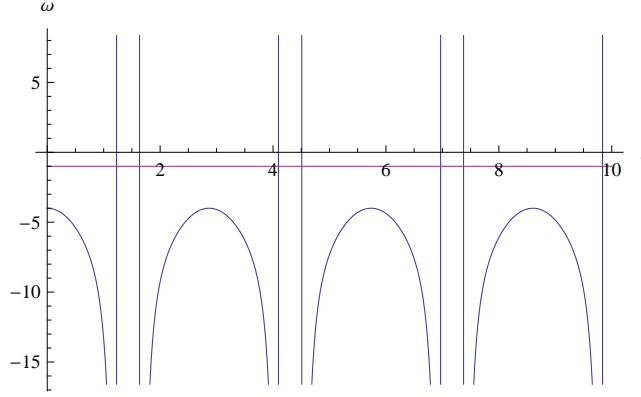}
  \caption{The EoS $\omega$ in Eq.~(\ref{FIG2_5}) as a function of $t$. 
Legend is the same as Fig.~\ref{fig:Example-1_5}.}
  \label{fig:Example-2_5}
\end{figure}

In Fig.~\ref{fig:Example-2_5}, we depict the cosmological evolution of 
EoS $\omega$ as a function of $t$ for $\wp\left(t,1,1\right)$.  
From Fig.~\ref{fig:Example-2_5}, we see that the universe always evolves 
within the phantom phase ($\omega < -1$). 
Consequently, this model is ruled out.

%
%
%
%

\subsection{MG -- XIV model}

We provide that the Hubble parameter is described as 
\begin{equation}
H=\ddot{\wp}(t)\,, 
\label{FIG3_1}
\end{equation} 
so that the scale factor can read as
\begin{equation}
a(t)=a_0\e^{\dot{\wp}(t)}\,.
\label{FIG3_2}
\end{equation}
{}From Eq.~(\ref{FIG3_1}) and the gravitational field equations (\ref{FE1}) 
and (\ref{FE2}), we have the parametric EoS
\begin{eqnarray}
p\Eqn{=}-24\wp(t)\sqrt{4\wp^3(t)-g_2\wp(t)-g_3}
-3\left(6\wp^2(t)-0.5g_2\right)^2\,,
\label{FIG3_3}\\
\rho\Eqn{=}3\left(6\wp^2(t)-0.5g_2\right)^2\,.
\label{FIG3_4}
\end{eqnarray}
Hence, 
the EoS parameter is written as 
\begin{equation}
\omega=-1-\frac{8\wp(t)\sqrt{4\wp^3(t)-g_2\wp(t)-g_3}}{\left(6\wp^2(t)-0.5g_2\right)^2}\,.
\label{FIG3_5}
\end{equation}
With Eqs.~(\ref{SF}) and (\ref{SIP}), we have 
\begin{eqnarray}
\phi\Eqn{=}2i\sqrt{6}\int\sqrt{\wp\left(t\right)\sqrt{4\wp^3\left(t\right)-g_2\wp\left(t\right)-g_3}}dt\,, \\
V\left(\phi\right)\Eqn{=}3\left(6\wp^2\left(t\right)-0.5g_2\right)^2+12\wp\left(t\right)\sqrt{4\wp^3\left(t\right)-g_2\wp\left(t\right)-g_3}\,.
\end{eqnarray}

\begin{figure}[h!]
  \centering
  \includegraphics[]{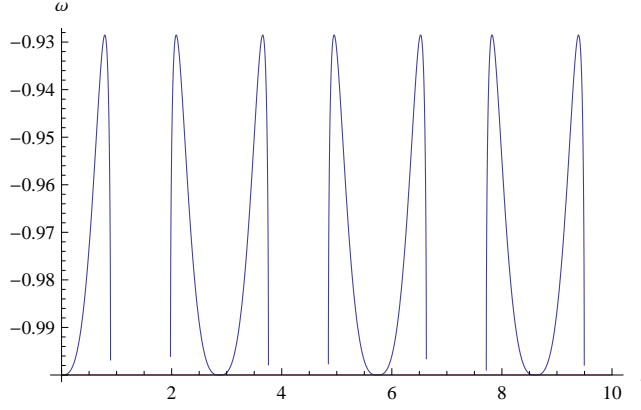}
  \caption{The EoS $\omega$ in Eq.~(\ref{FIG3_5}) as a function of $t$. 
Legend is the same as Fig.~\ref{fig:Example-1_5}.}
  \label{fig:Example-3_5}
\end{figure}

In Fig.~\ref{fig:Example-3_5}, we illustrate the cosmological evolution of 
EoS $\omega$ as a function of $t$ for $\wp\left(t,1,1\right)$. 
From Fig.~\ref{fig:Example-3_5}, we see that the universe always stays in 
the non-phantom (quintessence) phase ($\omega > -1$).

%
%

\subsection{MG -- XV model}

We take the form of the Hubble parameter as 
\begin{equation}
H=\dddot{\wp}(t)\,.
\label{FIG4_1}
\end{equation}
{}From this form, the corresponding scale factor is given by
\begin{equation}
a(t)=a_0\e^{\ddot{\wp}(t)}\,.
\label{FIG4_2}
\end{equation}
With Eq.~(\ref{FIG4_1}) and the gravitational field equations (\ref{SF}) and 
(\ref{SIP}), we obtain 
the parametric EoS 
\begin{eqnarray}
p\Eqn{=}-12\left(20\wp^3(t)-3g_2\wp(t)-2g_3\right)-432\wp^2(t)\left(4\wp^3(t)-g_2\wp(t)-g_3\right)\,,
\label{FIG4_3}\\
\rho\Eqn{=}432\wp(t)^2\left(4\wp^3(t)-g_2\wp(t)-g_3\right)\,.
\label{FIG4_4}
\end{eqnarray}
It follows from Eqs.~(\ref{FIG4_3}) and (\ref{FIG4_4}) that 
the parametric EoS is given by
\begin{equation}
\omega=-1-\frac{20\wp^3(t)-3g_2\wp(t)-2g_3}{36\wp(t)^2\left(4\wp^3(t)-g_2\wp(t)-g_3\right)}\,.
\label{FIG4_5}
\end{equation}
By using the formulae in Eqs.~(\ref{SF}) and (\ref{SIP}), 
we acquire 
\begin{eqnarray}
\phi\Eqn{=}2i\sqrt{3}\int\sqrt{20\wp^3\left(t\right)-3g_2\wp\left(t\right)-2g_3}dt\,, \\
V\left(\phi\right)\Eqn{=}6\left(20\wp^3(t)-3g_2\wp(t)-2g_3\right)+432\wp^2\left(t\right)\left(4\wp^3\left(t\right)-g_2\wp\left(t\right)-g_3\right)\,.
\end{eqnarray}

\begin{figure}[h!]
  \centering
  \includegraphics[]{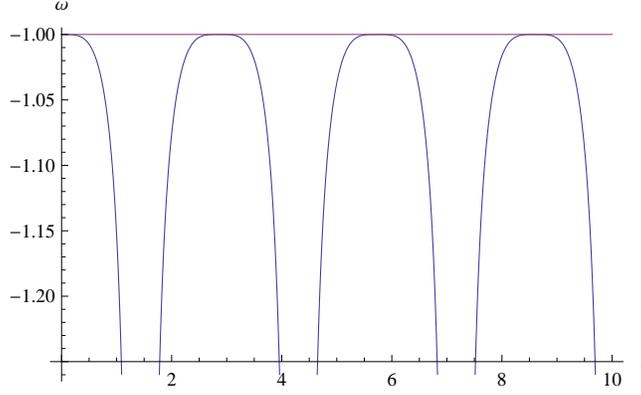}
  \caption{The EoS $\omega$ in Eq.~(\ref{FIG4_5}) as a function of $t$. 
Legend is the same as Fig.~\ref{fig:Example-1_5}.}
  \label{fig:Example-4_5}
\end{figure}

In Fig.~\ref{fig:Example-4_5}, we display the cosmological evolution of 
EoS $\omega$ as a function of $t$ for $\wp\left(t,1,1\right)$. 
From Fig.~\ref{fig:Example-4_5}, we understand that the universe always 
evolves within the phantom phase ($\omega < -1$). 
Thus, this model is ruled out. 

%
%

\subsection{MG -- XVI model}

We assume that the Hubble parameter is written as 
\begin{equation}
H=\ddddot{\wp}(t)\,.
\label{FIG5_1}
\end{equation}
In this case, the scale factor 
becomes
\begin{equation}
a(t)=a_0\e^{\dddot{\wp}(t)}\,.
\label{FIG5_2}
\end{equation}
Using Eq.~(\ref{FIG5_1}) and the gravitational field equations 
(\ref{FE1}) and (\ref{FE2}) yields 
\begin{eqnarray}
\hspace{-10mm}
p\Eqn{=}-36\left(20\wp^2(t)-g_2\right)\sqrt{4\wp^3(t)-g_2\wp(t)-g_3}-3\left(120\wp^3(t)-18g_2\wp(t)-12g_3\right)^2\,,
\label{FIG5_3}\\
\hspace{-10mm}
\rho\Eqn{=}3\left(120\wp^3(t)-18g_2\wp(t)-12g_3\right)^2\,.
\label{FIG5_4}
\end{eqnarray}
Hence, the parametric EoS is expressed as
\begin{equation}
\omega=-1-\frac{12\left(20\wp^2(t)-g_2\right)\sqrt{4\wp^3(t)-g_2\wp(t)-g_3}}{\left(120\wp^3(t)-18g_2\wp(t)-12g_3\right)^2}\,.
\label{FIG5_5}
\end{equation}
{}From Eqs.~(\ref{SF}) and (\ref{SIP}), we acquire 
\begin{eqnarray}
\hspace{-10mm}
\phi\Eqn{=}6i\int\sqrt{\left(20\wp^2(t)-g_2\right)\sqrt{4\wp^3\left(t\right)-g_2\wp\left(t\right)-g_2}}dt\,, \\
\hspace{-10mm}
V\Eqn{=}18\left(20\wp^2(t)-g_2\right)\sqrt{4\wp^3\left(t\right)-g_2\wp\left(t\right)-g_2}+3\left(120\wp^3\left(t\right)-18g_2\wp\left(t\right)
-12g_3\right)^2\,.
\end{eqnarray}

\begin{figure}[h!]
  \centering
  \includegraphics[]{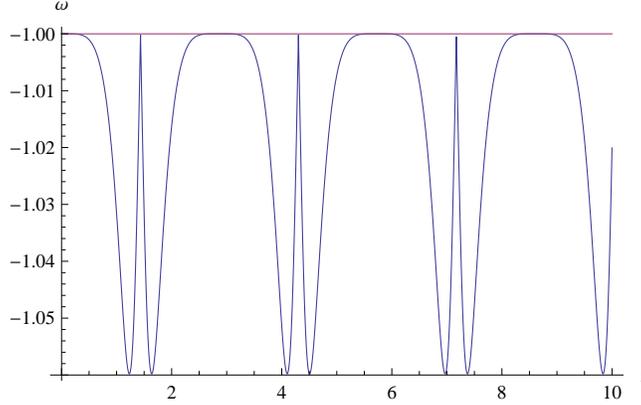}
  \caption{The EoS $\omega$ in Eq.~(\ref{FIG5_5}) as a function of $t$. 
Legend is the same as Fig.~\ref{fig:Example-1_5}.}
  \label{fig:Example-5_5}
\end{figure}

In Fig.~\ref{fig:Example-5_5}, we plot the cosmological evolution of 
EoS $\omega$ as a function of $t$ for $\wp\left(t,1,1\right)$. 
From Fig.~\ref{fig:Example-5_5}, we find that the universe always evolves 
within the phantom phase ($\omega < -1$). 
Therefore, this model is ruled out. 
In addition, we clearly see 
the two periodic oscillating evolution of $\omega$.

%
%

\subsection{MG -- XVII model}

We take the scale factor as 
\begin{equation}
a\left(t\right)=\wp(t)\,.
\label{FIG6_1}
\end{equation}
{}From this expression, the Hubble parameter becomes
\begin{equation}
H=\frac{\sqrt{4\wp^3\left(t\right)-g_2\wp\left(t\right)-g_3}}{\wp\left(t\right)}\,.
\label{FIG6_2}
\end{equation}
Equation (\ref{FIG6_2}) and the gravitational field equations (\ref{FE1}) and (\ref{FE2}) present the parametric EoS 
\begin{eqnarray}
p\Eqn{=}-\frac{16\wp^3\left(t\right)-2g_2\wp\left(t\right)-g_3}{\wp^2\left(t\right)}\,,\\
\rho\Eqn{=}3\frac{4\wp^3\left(t\right)-g_2\wp\left(t\right)-3g_3}{\wp^2\left(t\right)}\,.
\label{FIG6_4}
\end{eqnarray}
Furthermore, the EoS parameter is represented as 
\begin{equation}
\omega=-\frac{16\wp^3\left(t\right)-2g_2\wp\left(t\right)-g_3}{12\wp^3\left(t\right)-3g_2\wp\left(t\right)-3g_3}\,.
\label{FIG6_5}
\end{equation}
{}From Eqs.~(\ref{SF}) and (\ref{SIP}), we have 
\begin{eqnarray}
\phi\Eqn{=}i\int\frac{\sqrt{4\wp^3\left(t\right)+g_2\wp\left(t\right)+2g_3}}{\wp\left(t\right)}dt\,, \\
V\Eqn{=}\frac{28\wp^2\left(t\right)-5g_2\wp\left(t\right)-4g_3}{2\wp^2
\left(t\right)}\,. 
\end{eqnarray}

\begin{figure}[h!]
  \centering
  \includegraphics[]{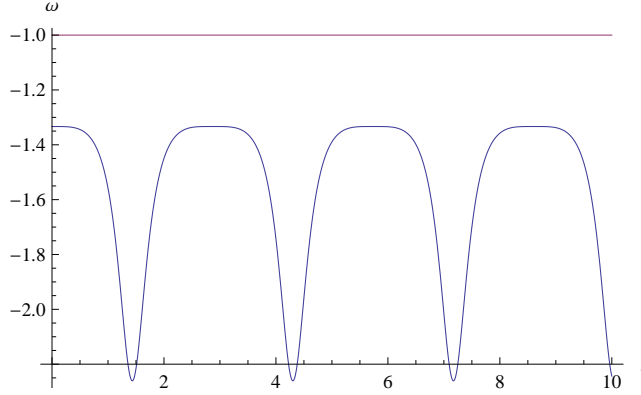}
  \caption{The EoS $\omega$ in Eq.~(\ref{FIG6_5}) as a function of $t$. 
Legend is the same as Fig.~\ref{fig:Example-1_5}.}
  \label{fig:Example-6_5}
\end{figure}

In Fig.~\ref{fig:Example-6_5}, we plot the cosmological evolution of 
EoS $\omega$ as a function of $t$ for $\wp\left(t,1,1\right)$. 
From Fig.~\ref{fig:Example-6_5}, we see that the universe always stays 
in the phantom phase ($\omega < -1$). 
Accordingly, this model is ruled out.

%
%

\subsection{MG -- XVIII model}

We describe the scale factor as 
\begin{equation}
a=\dot{\wp}(t)\,.
\label{FIG7_1}
\end{equation}
It follows from this description that the Hubble parameter is given by 
\begin{equation}
H=\frac{6\wp^2\left(t\right)-0.5g_2}{\sqrt{4\wp^3\left(t\right)-g_2\wp\left(t\right)-g_3}}\,.
\label{FIG7_2}
\end{equation}

\subsection{MG -- XIX model}

We consider that the scale factor is given by 
\begin{equation}
a=\ddot{\wp}(t)\,, 
\label{FIG8_1}
\end{equation}
so that the Hubble parameter can read 
\begin{equation}
H=\frac{12\wp\left(t\right)\sqrt{4\wp^3\left(t\right)-g_2\wp\left(t\right)-g_3}}{6\wp^2\left(t\right)-0.5g_2}\,.
\label{FIG8_2}
\end{equation}

\subsection{MG -- XX model}

We suppose the following form of the scale factor 
\begin{equation}
a=\dddot{\wp}(t)\,.
\label{FIG9_1}
\end{equation}
{}From this representation, 
the Hubble parameter is written by 
\begin{equation}
H=\frac{20\wp^3\left(t\right)-3g_2\wp\left(t\right)-2g_3}{2\wp\left(t\right)\sqrt{4\wp^3\left(t\right)-g_2\wp\left(t\right)-g_3}}\,.
\end{equation}

\subsection{MG -- XXXIII model}

We suppose that the scale factor takes the following form
\begin{equation}
a=\ddddot{\wp}(t)\,. 
\label{FIG10_1}
\end{equation}
It follows from this representation that the Hubble parameter is 
described by 
\begin{equation}
H=-\frac{3(g_2 - 20\wp^2\left(t\right))(g_3+g_2\wp^2\left(t\right) - 4\wp^3\left(t\right))}{2g_3 + 3g_2\wp\left(t\right) - 20\wp^3\left(t\right)}\,. 
\label{FIG9_2}
\end{equation}

It is important to mention that a quintom model 
(where there must exist both canonical and non-canonical scalar fields 
in order that the crossing of the phantom divide can occur) 
with single canonical scalar field cannot be reconstructed, 
in which the crossing of the phantom divide cannot occur~\cite{Zhao:2005vj}. 
Since in MG -- XVIII, XIX, XX and XXXIII models explained in 
Sections 4.7, 4.8, 4.9 and 4.10, respectively, 
if we reconstruct the corresponding single scalar field model, 
the crossing of the phantom divide can be realized. 
Thus, this implies that for these models, such a reconstruction 
is not physical but just mathematical procedure. 

The physical interpretation of our results is considered as follows. 
By reconstructing the expansion behavior of the universe, i.e., the Hubble 
parameter, with the Weierstrass $\wp(t)$-function, 
which has a periodic property, 
the periodic behavior of the EoS for the universe can be realized. 
Such a procedure can be applied to 
scenarios to avoid cosmological singularities and eventually a non-singular universe can be realized. 

It is also remarkable to note that in terms of all the numerical calculations 
in Figs.~\ref{fig:Example-1_5}--\ref{fig:Example-6_5}, 
the qualitative results do not strongly depend on the initial conditions and the model parameters such as the Weierstrass $\wp(t)$-function. 

Moreover, for all the models, 
the scale factor $a$ has to be positive and the Hubble parameter $H$ 
should be real. 
{}From Eqs.~(\ref{FIG1_2}), (\ref{FIG2_2}), (\ref{FIG3_2}), (\ref{FIG4_2}) 
and (\ref{FIG5_2}), $a$ is always positive. 
Also, it follows from Eqs.~(\ref{FIG6_1}), (\ref{FIG7_1}), (\ref{FIG8_1}), (\ref{FIG9_1}) and (\ref{FIG10_1}) that 
$a$ is written by $\wp(t)$ or its time derivatives. 
By using the Laurent expansion of $\wp(t)$ around $t=0$ 
as $\wp(t) = 1/t^2 + \left(g_2/20\right)t^2 + \left(g_3/28\right)t^4 + 
\mathcal{O}(t^6)$, 
we find 
$\dot{\wp}(t) = - 2/t^3 + \left(g_2/10\right)t + \left(g_3/7\right)t^3 + 
\mathcal{O}(t^5)$, 
$\ddot{\wp}(t) = 6/t^4 + g_2/10 + \left(3g_3/7\right)t^2 + 
\mathcal{O}(t^4)$, 
$\dddot{\wp}(t) = -24/t^5 + \left(6g_3/7\right)t + 
\mathcal{O}(t^3)$, 
and
$\ddddot{\wp}(t) = 120/t^5 + 6g_3/7 + 
\mathcal{O}(t^2)$. 
Thus, for the cases of $a=\dot{\wp}(t)$ in Eq.~(\ref{FIG7_1}) and 
$a=\dddot{\wp}(t)$ in Eq.~(\ref{FIG9_1}), around $t=0$, 
the expression of $a$ is not well defined because 
the value of $a$ would become negative around $t=0$. 
In this sense, around $t=0$, 
the MG -- XVIII and MG -- XX models cannot be used. 
On the other hand. 
since the argument of the Weierstrass functions used 
in this paper is the cosmic time $t$, which is real, all the values of the Weierstrass functions are real. In all the models, $H$ is described by 
the Weierstrass functions and hence $H$ would be real. 

Furthermore, we state that 
if the evolution of $\omega$ is displayed as a function of the redshift 
$z \equiv a_{\mathrm{p}}/a - 1$ with $a_{\mathrm{p}}$ the present value of $a$, 
the direction of the cosmic time $t$ to go by becomes opposite to that 
for $\omega$ to be shown as a function of $t$ 
in Figs.~\ref{fig:Example-1_5}--\ref{fig:Example-6_5}. 
It would be considered that the main qualitative 
difference is only this point and hence other cosmological consequences 
do not change. 

In addition, 
it is interesting to note that 
the models examined in this work 
may solve, not only the dark energy paradigm, but also the horizon and flatness problems. 
In other words, 
these models may produce a kind of inflationary epoch or perhaps a contraction phase as in the ekpyrotic scenario. 
In order for inflation or the ekpyrotic scenario to be realized realistically, 
it is necessary to consider the way of connect the inflationary stage and 
the dark energy dominated stage, i.e., the realization of the reheating stage. 
This is a crucial point for these models to be realistic inflation models 
or the ekpyrotic scenario. 

Moreover, 
it should be cautioned that 
in Figs.~\ref{fig:Example-2_5}--\ref{fig:Example-4_5}, 
apparently it looks the EoS $\omega$ diverges. 
However, the reason is just a way of plotting and 
therefore there is no divergence in terms of the EoS in all of the figures.  

Finally, we mention the issue of existence of a ghost, namely, the instability 
for the case of the crossing of the phantom divide. 
For a scalar field theory, it is known that if 
the crossing of the phantom divide happens in the FLRW universe, 
there would appear a ghost. 
In Sec.~IV, we have presented the interpretations of our models 
in the framework of a scalar field theory. 
However, there still exists the possibility that our models could be regarded 
as a more complicated theory, e.g., which is described by the non-linear 
kinetic terms such as $k$-essence models~\cite{
Chiba:1999ka, ArmendarizPicon:1999rj, Garriga:1999vw, ArmendarizPicon:2000dh, 
ArmendarizPicon:2000ah, dePutter:2007ny} 
and the Galileon models~\cite{
Nicolis:2008in, Deffayet:2009wt, Deffayet:2009mn, Deffayet:2010zh, 
Shirai:2012iw}. 
These investigations might be one of the interesting future works in our 
approach to the periodic and quasi-periodic generalizations of the Chaplygin gas type models. 

\section{Conclusions}

In the present paper, we have reconstructed 
periodic and quasi-periodic 
generalizations of the Chaplygin gas type models by using 
the Weierstrass $\wp(t)$, $\sigma(t)$ and 
$\zeta(t)$-elliptic functions.  
We have explored the cosmological evolution of the EoS for dark energy in FLRW spacetime. 
In particular, we have shown that 
by using the degenerate properties of the elliptic functions, 
the MCG models can be recovered. 
This is one of the most important cosmological ingredients obtained in this 
paper. This result implies that 
by plugging the reconstruction method of the expansion history of the universe 
with the Weierstrass elliptic functions, we can acquire the MCG models, 
which have a potential to reveal the properties of both dark energy and dark matter. In other words, the procedure demonstrated in this paper can lead to a preferable cosmological model by starting with the mathematical special functions. 

It is meaningful to summarize the following significant points. 
(a) The Weierstrass functions have two periods $m_1$ and $m_2$. 
This is their essential property. 
(b) Our periodic and quasi-periodic models given in Sec.~III are periodic or quasi-periodic in terms of the energy density $\rho$.  
(c) Furthermore, 
if the special values of the periods $m_1$ and $m_2$ are infinite, 
these models in Sec.~III are 
transformed into or have limits of the Chaplygin gas models, 
which can be seen from the formulae in (3.1). 
Thus, in this sense to reconstruct these models is considered to be two periodic or quasi periodical generalizations of the CG models. 
This is the justification of application of the Weierstrass functions 
to cosmological models. 
(d) Our models presented in Sec.~IV are periodic on the cosmic time $t$ (which is here dimensionless) contrast to the models in Sec.~III which are periodic/quasiperiodic on $\rho$. 
(e) Moreover, 
our periodic/quasiperiodic models in Sec.~III are singular at $\rho=0$ 
and those in Sec.~IV at $t=0$. 

It is also important to emphasize that 
the cosmological advantage to acquire the periodic evolution of the 
universe is to realize scenarios to avoid a Big Bang singularity, the finite-time future singularities and a Big Crunch singularity. By applying the obtained results and the discussed procedure to scenarios for the avoidance of 
singularities, e.g., the cyclic universe, the ekpyrotic scenario and 
the bouncing universe, we can find a non-singular cosmology. 

Furthermore, 
it has explicitly been illustrated that there are three type models with 
realizing the cosmological evolution of the EoS for dark energy
: (i) The universe always stays in the non-phantom (quintessence) phase 
(the MG -- XIV model). 
(ii) The universe always evolves within the phantom phase (the MG -- XII, MG -- XIII, MG -- XV, MG -- XVI and MG -- XVII models). 
(iii) The crossing of the phantom divide can be realized (the MG -- XVIII, MG -- XIX, MG -- XX and MG -- XXXIII models). 
If the universe always stays in the phantom phase, 
it is impossible to describe the whole evolution history of the universe, 
i.e., the decelerated phases such as the radiation and matter dominated 
stages. Thus, these models are ruled out. 
In addition, the corresponding description of 
a canonical scalar field model to the models in the above (iii) category 
is not physical but just mathematical because in the light of 
quintom model~\cite{Zhao:2005vj}, 
it is impossible for a single canonical scalar field to realize 
the crossing of the phantom divide. 

The scalar fields and the corresponding potentials have been analyzed for 
different types of MG models mentioned above. The EoS parameters 
have been derived and their natures have also been illustrated 
graphically during the evolution of the universe. 

On the other hand, 
at the current stage the cosmological constant is consistent with 
the observational data~\cite{Komatsu:2010fb}, but 
there still exists the possibility of dynamical dark energy, which is, e.g., 
described by a scalar field or a fluid. 
Thus, when we acquire the results of data analysis of 
the more precise future experiments like 
by the PLANCK satellite~\cite{Planck-HP}, 
it is strongly expected that 
the investigations on the phase of the universe, i.e., 
the non-phantom or phantom phase, or the crossing of the phantom divide
can become more meaningful. 
In the future work, it would be interesting that 
whether the above models may be suitable candidates for dark energy 
is tested by coming observational data fittings.

\section*{Acknowledgments}
K.B. would like to sincerely 
acknowledge very kind and warm hospitality 
at Eurasian National University very much, where this work has been completed. 
In addition, he is grateful to National Center for Theoretical 
Sciences and National Tsing Hua University very much for the very kind and 
warm hospitality, where the revision of this work was executed.

\end{document}